\documentclass[twocol]{ametsocV6.1}

\title{Extended Analysis of the Effects of the Sumatra’s Topography on Downstream Low-level Vortex Development over the Indian Ocean}

\authors{Paul E.\ Ciesielski\correspondingauthor{Paul E.\ Ciesielski, ciesiels@colostate.edu} and 
Richard H.\ Johnson
}
\affiliation{{Department
of Atmospheric Science, 1371 Campus Delivery, Colorado State
University, Fort Collins, CO 80523}
}

\abstract{ Fine et al.\! (2016, hereafter F16) investigated the potential role of Sumatra Island, as well as the Malay Peninsula and Java, in creating terrain-induced circulations over the Indian Ocean (IO) that subsequently develop into tropical cyclones (TCs). Applying sophisticated vortex tracking software to 2.5 yrs of model analyses, F16 found four regions downstream of topographic features in the Maritime Continent to be prolific generators of low-level cyclonic vortices (123/yr). Vortices shed from these “hotspot” regions contributed to 25\% of all TCs occurring in the IO basin during that period. This present study extends the limited analyses of F16 by applying a similar approach to 10 yrs (2008-2017) of ERA5 analyses. While the 2.5-yr period which F16 studied was slightly ($\sim$8\%) more active in terms of vortex production than the 10-yr period, in general F16’s findings are representative of the longer-term record with $\sim$70\% ($\sim$80\%) of all vortices (shed vortices) occurring with easterly low-level (925 hPa) flow over the “hotspot” regions.
Additional analysis of the 10-yr record found that vortex counts are highest near MJO phase 1 when low-level easterlies are strongest over the maritime continent region. A secondary peak in vortex (and shed vortex) counts occurs during MJO phase 4 when low-level westerlies were present near the equator west of Sumatra. This suggests that low-level westerly surges on the equator impinging on Sumatra associated with the MJO contribute to an increase in wake vortex development. While the frequency of vortex genesis over the four “hotspot” regions is strongly tied to the annual cycle of winds, periods with anomalous zonal flow are shown to impact vortex counts. This is most apparent in the region off the southern tip of Sumatra (SS) where vortex counts were 4 times higher during periods of anomalous easterlies compared to periods with anomalous westerlies. While positive (negative) El-Nino-Southern Oscillation (ENSO) and Indian-Ocean Dipole (IOD) conditions drive anomalous easterlies (westerlies), the relationship between vortex count and these large-scale indices is weak (r < 0.1) in most regions. Only in region SS did the large-scale conditions associated with these indices appear to impact vortex formation frequency in the sense that positive ENSO and IOD conditions drive anomalous easterlies which result in higher vortex formation rates.
 }

\begin{document}

\maketitle

\section{Introduction}

Under certain atmospheric conditions, stratified flow can be blocked by topographic features resulting in downstream formation of wake vortices \citep{smolarkiewicz1989low, rotunno1991further}. \cite{kuettner1967equatorial, kuettner1989easterly}  was the first to propose that Sumatra may serve as a generator of wake vortices that subsequently develop into tropical cyclones (TCs). Easterly flow blocked by Sumatra’s narrow mountain range, which exceeds 3000m in places and straddles the equator from 6$^\circ$N to 6$^\circ$S, can result in the formation of wake cyclonic vortices over the India Ocean (IO). Due to the paucity of observations over the IO, Kuettner’s observations of these wake vortices have received little attention until recently.

As a result of limitations in our basic understanding of, and ability to predict the intraseasonal oscillation, particularly its initiation, a field campaign known as the Dynamics of the Madden-Julian Oscillation (DYNAMO) was conducted in the Indian Ocean (IO) region during the period October 2011–March 2012 \citep{YoneyamaEA2013}. This experiment provided researchers with unprecedented atmospheric and oceanic datasets over a data-sparse region of our planet including observations from upper-air soundings from ships and islands \citep{CiesielskiEA2014}, radars, aircraft, and satellites. During DYNAMO a wake vortex initiated off the northern tip of Sumatra in November 2011 as low-level easterlies associated with an approaching MJO convective envelope \citep{Johnson+Ciesielski2013, GottschalckEA2013} flow impinged upon the island. This vortex drifted slowly westward for four days developing into a TC which caused significant damage and deaths in Sri Lanka (http:// www.webcitation.org/63Vnhm01u). This storm and other TCs, which formed from wake vortices during DYNAMO, resulted in a renewed an interest in this topic.

\cite{Fine2016role} (hereafter F16) investigated the potential role of Sumatra Island, as well as the adjacent topography of the Malay Peninsula and Java (Fig.\ 1), in producing terrain-induced circulations over the Indian Ocean that subsequently develop into TCs. Applying sophisticated vortex tracking software (Hodges 1995, 1999) to high-resolution (0.25°) model analyses, F16 found that these topographic features are prolific generators of low-level cyclonic vortices which contributed to 25\% of the TCs occurring in the IO basin during their study period. F16’s findings were based on a limited period (2.5 yr) in which high-resolution ECMWF analyses were available (YOTC analyses from May 2008 to April 2010) and 6 months of a special DYNAMO operational analyses (from October 2011 to March 2012). 

With the recent availability of the ECMWF Reanalysis 5th Generation (or ERA5) high-resolution dataset for the years 1979 to present, the opportunity to extend the F16 analyses to additional years became possible. The goal of this study is to determine the representativeness of F16’s findings compared to a longer period (2008-2017) and a different, and presumably improved, reanalysis dataset. Utilizing this longer analysis period, we investigate variability of vortex formation and its relationship to various large-scale signals such as the MJO, IOD, and ENSO signals. Finally, we consider some long-term IO TC statistics to examine the relationship of TCs to wind regimes and large-scale indices.
       
\begin{figure*}[thbp]
\centerline{\includegraphics[width=6in]{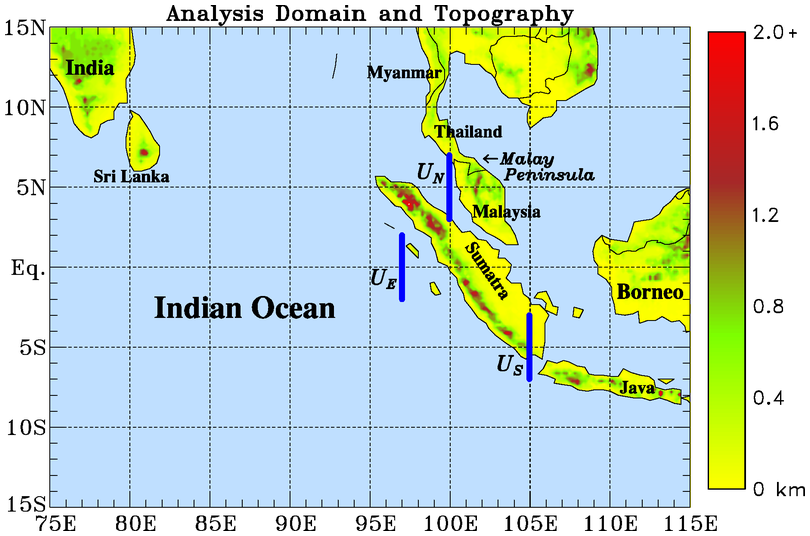}}

\protect\caption{Analysis domain for study and topography of the region (elevation scale to right in km). Blue lines at: 3$^\circ$-7$^\circ$N,100$^\circ$E, 2$^\circ$N-2$^\circ$S, 97$^\circ$E, and 3$^\circ$-7$^\circ$S, 105$^\circ$E represent averaging areas for zonal winds corresponding to these line segments ($U_N$, $U_E,$ and $U_S$, respectively) to delineate the various flow regimes in the area which affect lee vortex production.
}\label{f1}
\end{figure*}

\section{Data and Methodology}

\subsection{Datasets}

ECMWF Reanalysis 5th Generation (or ERA5), which replaces the ERA-Interim reanalysis, is based on 4D-Var data assimilation using Cycle 41r2 of the Integrated Forecasting System (IFS),
which was operational at ECMWF in 2016. A detailed description of the ERA5 configuration, which benefits from a decade of developments in model physics, core dynamics, and data assimilation relative to ERA-Interim, can be found in \cite{hersbach2020}. For this study we used both vorticity and zonal winds from ERA5.

The Madden-Julian Oscillation (MJO) is the major fluctuation in tropical weather on weekly to monthly timescales \citep{Madden1971}. The MJO can be characterized as an eastward moving 'pulse' of cloud and rainfall near the equator that typically recurs every 30 to 60 days. The location and strength of the MJO are given by the Real-time Multivariate MJO (RMM) index developed by \cite{Wheeler+Hendon2004}. This index is based on a pair of empirical orthogonal functions (EOFs) of the combined fields of near-equatorially averaged 850-hPa zonal wind, 200-hPa zonal wind, and satellite-observed outgoing longwave radiation (OLR) data. Daily values of the RMM index provide the amplitude and phase of the MJO. For this study, classification of the vortices by MJO phase was only done when the MJO signal had a significantly robust amplitude defined here as when the RMM amplitude was > 1.

To represent the variability of the El Niño/Southern Oscillation (ENSO) we use the Multivariate ENSO Index Version 2 (MEI.v2) which combines both oceanic and atmospheric variables into a single assessment of the state of ENSO \citep{Wolter2011nino, Zhang2019towards}. Positive (negative) values of this index imply warm, El Niño (cool, La Niña) conditions across the east-central equatorial Pacific. The MEI.v2 is available on a monthly basis.

Variability in the IO region is often characterized by changes in the east-west sea surface gradient across the IO basin referred as the Indian Ocean Dipole (IOD) mode \citep{saji1999dipole}. The IOD is commonly measured by the Dipole Mode Index (DMI), that is, the difference between sea surface temperature (SST) anomalies between the western (50$^\circ$-70$^\circ$E) and eastern (90$^\circ$-110$^\circ$E) tropical (10$^\circ$S-10$^\circ$N) IO. A positive (negative) IOD period is characterized by cooler (warmer) than average water in the tropical eastern Indian Ocean and warmer (cooler) than average water in the tropical western Indian Ocean.

Information on IO tropical cyclone tracks and intensity was obtained from the Joint Typhoon Warning Center (JTWC) best-track archive which goes back to 1945 \citep{chuEA2002}. For this study TC data were used from 1980 to 2017, although \cite{chuEA2002} state that the years after 1984 have the best data quality. The best-track data contains the storm center locations and intensities (i.e., the maximum 1-minute mean sustained 10-meter wind speed) at six-hour intervals. Storms are only considered here that (1) reach TC intensity (i.e., maximum sustained winds of 35 knots or 39 mph) and (2) formed over the IO (i.e., extending from Africa to 105$^\circ$E). Storms with missing wind speeds, which are most prevalent in the earlier years of the archive, were not considered.

\subsection{Vortex tracking}

As in F16, identification and tracking of low-level lee vortices was carried out based on the relative vorticity field using the objective feature tracking code of \cite{hodges1995feature, hodges1999adaptive}. To facilitate the use of this software, relative vorticity at 6 h and 0.25$^\circ$ horizontal resolution from ERA5 analyses were vertically averaged using data at 850, 875, 900, 925 hPa from 50$^\circ$-110$^\circ$E, 20$^\circ$N-20$^\circ$S. The F16 analyses had only three vertical levels in the 850 to 925 hPa layer. After smoothing the vorticity field to retain spatial scales greater than 450 km, cyclonic vortex features were tracked if they maintained an amplitude greater than 1.0 $\times$ 10$^{-5}$ s$^{-1}$ for longer than 2 days.

To focus on wake vortices with the potential to develop into TCs, we restrict our analyses to cases where the vortex moved westward over the Indian Ocean. As in F16 these shed vortices are defined as those with 1) a final location over the Indian Ocean that is > 500 km from Sumatra, and either 2) a final minus initial displacement from Sumatra > 250 km, or 3) their average speed away from Sumatra is > 0.5 ms$^{-1}$. The first condition ensures that the shed vortex at the end of its track is some critical distance from Sumatra while conditions 2 or 3 guarantee that the vortex is moving away from its generating landmass.

Application of the tracking code to ten years of ERA5 data yields information on both the tracks and locations of cyclonic lee vortices. Because of seasonal changes in the flow regime, genesis locations are shown in Fig.\ 2 for the boreal cold season (November to April) and the boreal warm season (May to October). Overall, these analyses based on 10 years of statistics are remarkably similar to those shown in F16. For example, during the boreal winter season (Fig.\! 2a), northeasterly flow across the Malay Peninsula and northern tip of Sumatra result in a high frequency of lee vortices being formed downstream of these topographic features. Two boxes shown in this figure (covering the same areas as in F16) are defined to help quantify the relationship between the winds and vortex generation over these regions in subsequent analyses. These two regions are referred to a SN (Sumatra North) and MP (Malay Peninsula). Other hotspots for lee vortex generation are also seen over Sri Lanka and the west coast of India but are not considered further in this study. Between the equator and 10$^{\circ}$S, westerly flow predominates such that production of lee vortices over the IO west of the any topographic features is much less common in this area.

During the boreal summer monsoon (Fig.\ 2b), southeasterly flow prevails across the southern tip of Sumatra and Java resulting in a high frequency of vortex production in the lee of these topographical features. As in F16, analysis boxes are defined as SS (Sumatra South) and JV (Java) to highlight these lee vortex hotspots for further analysis. Westerly flow dominates the NH IO region during this period resulting in several lee vortex hotspots. However, these areas are not considered further as any vortices shed from these areas would move eastward and likely not contribute to any TCs.

To better understand the relationship of different flow regimes on the production of lee vortices, three computational lines shown in Fig.\! 1 are defined: (1) 3$^\circ$-7$^\circ$N, 100$^\circ$E, (2) 2$^\circ$N-2$^\circ$S, 97$^\circ$E, and (3) 3$^\circ$-7$^\circ$S, 105$^\circ$E. Averages of zonal winds in these 3 areas, referred to as $U_N$, $U_E$, and $U_S$, respectively, are considered representative of the flow in the adjacent analysis boxes and over the equator west of Sumatra. In regressing relative vorticity in the analysis boxes onto zonal winds, $U_N $ and $U_S$, F16 found that correlations maximize around 0.8 in the 950-850 hPa layer in the north and around 0.6 in the 950-1000 hPa layer to the south. Such correlations are expected considering flow blocking effects of topography and the average height of 1-2 km in northern Sumatra and slightly lower in the south. Thus, for future analyses involving the winds in this study, data from the 925 hPa level are used.

\begin{table*}[thbp]

\caption{Vortex formation and shedding statistics for four regions (SN, MP, SS, and JV) and based on different analyses (YOTC/DYNAMO used in F16 and ERA5) and analysis periods (2.5 yr period examined in F16 and 2008-2017 period). Numbers in parentheses represent annual rates.
}
\label{t1}
\begin{center}
\begin{tabular}{lccc}
\hline\hline
Region (Analysis, Years) & Total Vortices & \# Shed Vortices & Percent Shed  \\

\hline
SN (YOTC/DYN, 2.5 yr) & 73 (29.2) & 43 (17.2)  & 59\%  \\
 SN (ERA5, 2.5 yr) & 69 (27.6) & 40 (16.0) &  58\% \\
SN (ERA5, 10 yr) & 248 (24.8) &  143 (14.3) &  58\% \\
\hline
MP (YOTC/DYN, 2.5 yr) & 85 (34.0) & 15 (6.0) & 18\%  \\
 MP (ERA5, 2.5 yr) & 75 (30.0)  & 21 (8.4) & 28\% \\
MP (ERA5, 10 yr) & 295 (29.5) & 90 (9.0) & 31\% \\
\hline
SS (YOTC/DYN, 2.5 yr) &74 (29.6) & 25 (10.0)   & 34\%  \\
SS (ERA5, 2.5 yr) & 69 (27.6) & 20 (8.0) & 29\% \\
SS (ERA5, 10 yr) & 272 (27.2) & 80 (8.0) & 29\% \\
\hline
JV (YOTC/DYN, 2.5 yr) & 77 (30.8) & 20 (8.0)  & 26\%  \\
JV (ERA5, 2.5 yr) & 96 (38.4)  & 49 (19.6) & 51\%  \\
JV (ERA5, 10 yr) & 326 (32.6) & 185 (18.5) & 57\% \\
\hline
\hline

\end{tabular}
\end{center}
\end{table*}

\section{Results}

\subsection{Comparison to Fine et al. 2016}

Vortex statistics for the four vortex production hotspots identified in Fig.\ 2 are listed in Table 1 for the YOTC/DYN (YD) period. For each region, we compare the number of total and shed vortices from F16 to that from ERA5 (referred hereafter as EYD). In general, the statistics for F16 and EYD agree reasonably well except for the JV region in which EYD has ~25\% more total vortices and more than double the number of shed vortices. Considering all four hotspot regions, Table 2 show that the total the number of vortices formed is identical between these analyses (i.e., 309) but the percentage shed is higher in EYD (42\% versus 33\%) primarily due to the JV region with its high shed frequency.

\begin{figure}[thbp]
\centerline{\includegraphics[width=3in]{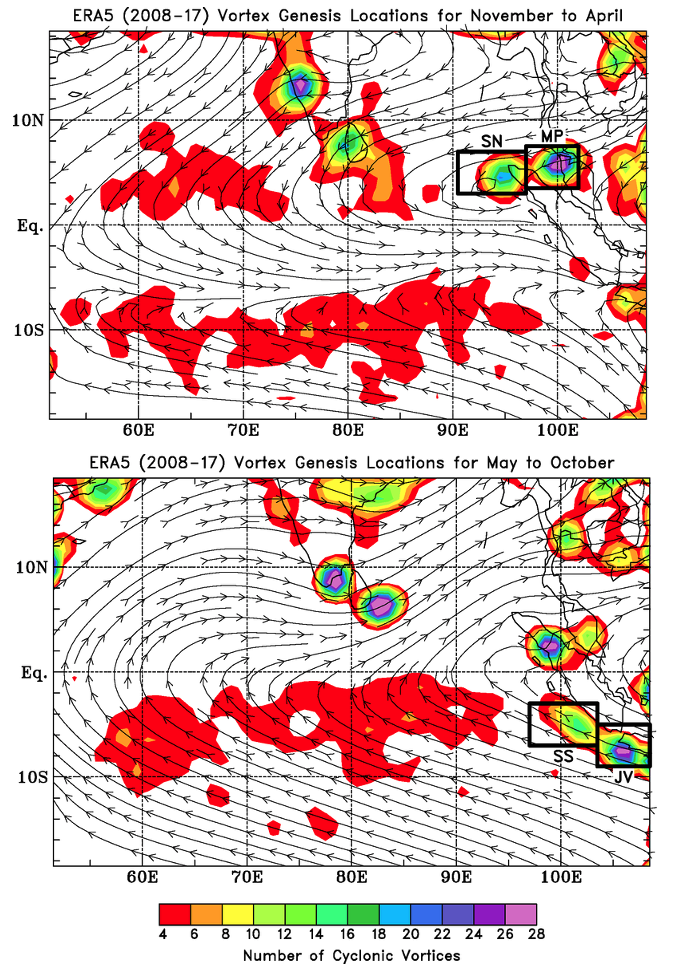}}

\protect\caption{Cyclonic vortex genesis locations on a 1$^\circ$ grid for (a) November to April and (b) May to October periods based on 10-yr (2008-2017) analyses of ERA5 data. Streamlines depict the 925- 850 hPa layer flows. Boxes highlight vortex production hotspots which are used for further analyses in this study.
}\label{f2}
\end{figure}

F16 showed a strong relationship between the winds in the vicinity of the topographic features and the production of lee vortices. In their study they found that about 80\% (85\%) of vortices (shed vortices) occurred in the presence of low-level easterly flow (F16 Table 1). Similar statistics are present for the EYD (not shown). Comparison of low-level winds between the F16 and the EYD analysis show little difference (Fig.\  3). For example, the temporal correlation of daily-averaged $U_S$ winds between these analyses is 0.99 and a difference in means is less than 0.2 m s$^{-1}$, with F16 having slightly stronger easterlies (or weaker westerlies) throughout the year. With this similarity in winds between the two analyses, it difficult to understand why the percentage of shed vortices in the JV region is nearly double in EYD.

F16 identified 13 TCs which developed from shed lee vortices in their 2.5 yr analyses. These TCs were tracked primarily by using the vorticity field but in 4 cases the shed vortex amplitude fell below the threshold limit of 1.0 $\times$ 10$^{-5}$ s$^{-1}$ such that continuity of the tracked vortex into a TC was established by also tracking the vortex as a depression in the height field. Using ERA5 we were able to identify these 13 TCs with similar tracking issues as in F16.

Overall, the good agreement between F16 and that from the ERA5 for the YD period gives us confidence that the ERA5 has reproduced the salient feature of vortex formation, shedding, and development into TCs found in F16. Since ERA5 has improved model physics and data assimilation over the analysis used in F16, it seems reasonable to assume that the notable differences we did find (e.g., higher vortex production and shedding over JV region using ERA5) are likely better represented in ERA5. We now present statistics for the longer 10-yr record to determine if the findings of F16 are representative of this longer-term period.

\begin{table}[thbp]

\caption{
Vortex formation and shedding statistics for four regions (SN, MP, SS, and JV) combined based on different analyses (YOTC/DYAMO used in F16 and ERA5) and analysis periods (2.5 yr period examined in F16 and 2008-2017 period). Numbers in parentheses represent the annual rate (i.e., vortices/year).
}
\label{t2}
\begin{center}
\begin{tabular}{lccc}
\hline\hline
Analysis, Years & Total Vortices & \# Shed Vortices & Percent Shed  \\

\hline
F16, 2.5 yr   & 309 (123.6) & 103 (41.2) & 33\% \\
\hline
EYD, 2.5 yr & 309 (123.6) & 130 (52.0) & 42\% \\
\hline
ERA5 10yr & 1141 (114.1) & 498 (49.8) & 44\% \\
\hline
\hline
\end{tabular}
\end{center}
\end{table}

\begin{table}[t]

\caption{
Tropical cyclone counts over the NH and SH of the Indian Ocean for the YD, 2008-17, and 1980-2017 periods. Numbers in parentheses represent the annual rates (i.e., TCs/year).
}
\label{t3}
\begin{center}
\begin{tabular}{lccc}
\hline\hline
Period & NH IO & \# SH IO & NH+SH IO  \\

\hline
YD (2.5 yr)  & 15 (6.0) & 34 (13.6) & 49 (19.6) \\
\hline
2008-17 (10 yr) & 51 (5.1) & 105 (10.5) & 156 (15.6) \\
\hline
1980-2017 (38 yrs) & 181 (4.8) & 434 (11.4) &  615 (16.2) \\
\hline
\hline
\end{tabular}
\end{center}
\end{table}

\subsection{Vortex statistics based on 10 years of ERA5}

To place the F16 analyses in the context of a longer period, Tables 1 and 2 include the vortex formation and shed statistics for the 2008-2017 period. To facilitate this comparison between periods of different length, the numbers in parentheses list the annual rate (i.e., vortex count divided by the number of years). Comparing the ERA5 results for the two periods shows that YD-period was characterized with a higher frequency of vortex formation in all four regions (Table 1). Considering the regions collectively (Table 2), the YD-period showed an ~8\% rate increase in vortex formation and an $\sim$4\% increase in shed vortex numbers compared to the 10-yr mean. Further examination of fields at 925 hPa over these regions revealed that the YD-period had slightly stronger (0.3 m s$^{-1}$) easterly winds and higher cyclonic relative vorticity (6$\times$10$^{-7}$s$^{-1}$) compared to the 10-yr period. It’s worth noting that the TC frequency over the IO was ~20\% higher in YD-period compared to 10-yr mean (Table 3). In summary, the YD-period was slightly more active than the 10-yr mean in terms of vortex formation, occurrence of shed vortices, and TCs, a likely consequence of an enhanced easterly component of the flow during the YD period. Having established a context for the YD-period, we now consider some additional characteristics of lee vortex production in the 10-yr period.

\begin{figure}[thbp]
\centerline{\includegraphics[width=3in]{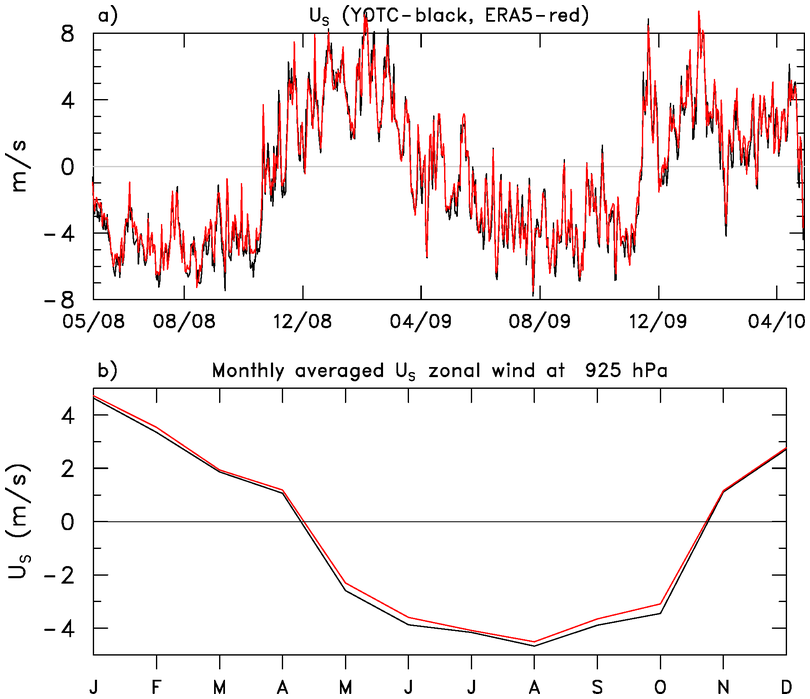}}

\protect\caption{ a) Time series of daily-averaged winds over southern Sumatra ($U_S$) for the YOTC period (May 2008 to April 2009) from F16 (black) and ERA5 (red), b) similar to a) expect annual cycle of monthly averaged $U_S$ winds.
}\label{f3}
\end{figure}

\begin{figure}[thbp]
\centerline{\includegraphics[width=3in]{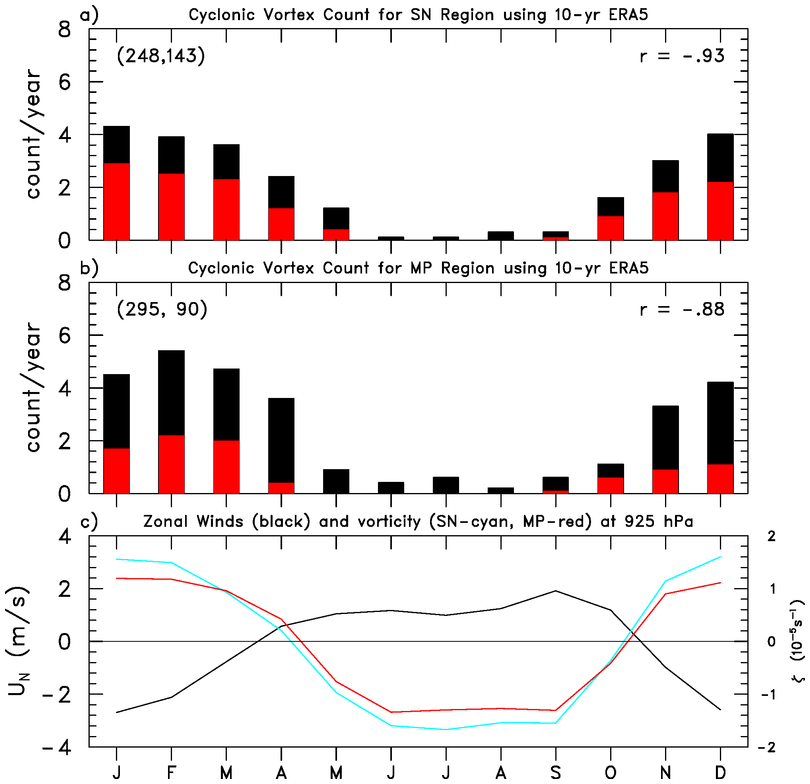}}

\protect\caption{
Monthly frequency of terrain-induced cyclonic vortices (top of black bar) and those which shed (top of red bar) for (a) SN region and (b) MP region. (c) Monthly mean zonal winds (m s$^{-1}$, scale to left) at 925 hPa along 100$^\circ$E ($U_N$ line segment shown in Fig.\ 1), and mean monthly vorticity (10$^{-5}$s$^{-1}$, scale to right) at 925 hPa over SN (cyan) and MP (red). The numbers in the parentheses indicate the total number of vortices formed and those that shed, respectively, over each region. Numbers in upper-right side of panels represent the correlation between the monthly vortex frequency and the monthly-averaged zonal winds over $U_N$.
}\label{f4}
\end{figure}

\begin{figure}[thbp]
\centerline{\includegraphics[width=3in]{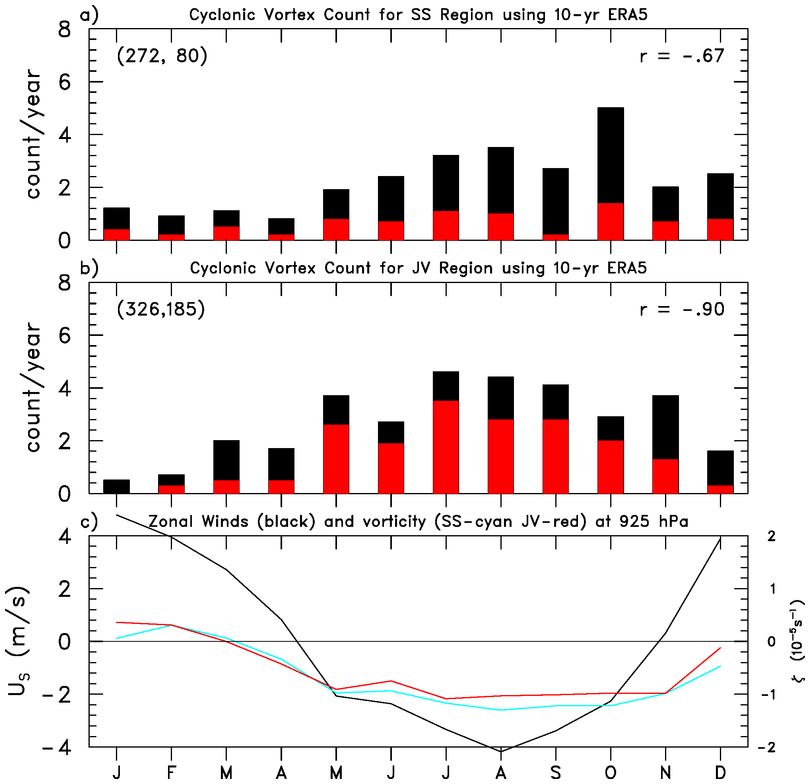}}

\protect\caption{
Monthly frequency of terrain-induced cyclonic vortices (top of black bar) and those which shed (top of red bar) for (a) SS region and (b) JV region. (c) Monthly mean zonal winds (m s$^{-1}$, scale to left) at 925 hPa along 105$^\circ$E ($U_S$ line segment shown in Fig.\ 1), and mean monthly vorticity (10$^{-5}$s$^{-1}$, scale to right) at 925 hPa over SS (cyan) and JV (red). Numbers in parentheses indicate the total number of vortices formed and those that shed, respectively, over each region. Numbers in upper-right side of panels represent the correlation between the monthly vortex frequency and the monthly-averaged zonal winds over $U_S$.
}\label{f5}
\end{figure}

\begin{figure}[thbp]
\centerline{\includegraphics[width=3in]{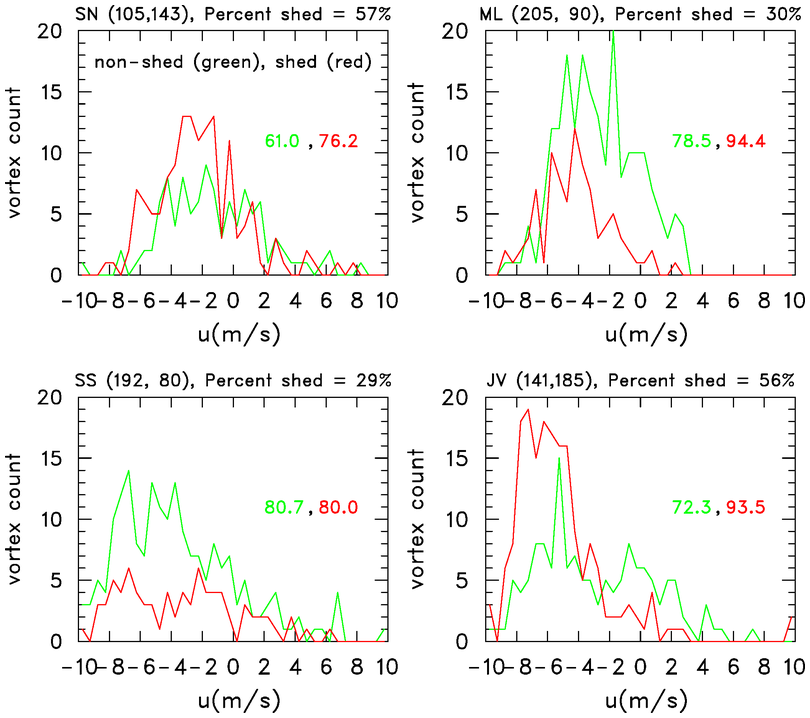}}

\protect\caption{
Vortex count as a function of zonal wind (m s$^{-1}$) for non-shed (green) and shed vortex (red) cases for the four regions of interest in this study. Numbers in parentheses in top labels represent the number of non-shed and shed vortex cases, respectively. Within each panel the numbers give the percentage of cases in which the vortex was initialized under easterly flow conditions (non-shed – green, shed – red).
}\label{f6}
\end{figure}

\begin{table}[t]

\caption{
Zonal wind (m s$^{-1}$) statistics for regions of interest. From left to right: Mean zonal wind for region ($\bar{U}$) based on 2008-17 period, average zonal wind associated with vortex formation ($U\!V\!F$), difference between $U\!V\!F$ and $\bar{U}$, $U\!V\!F$ for non-shed vortices, $U\!V\!F$ for shed vortices.
}
\label{t4}
\begin{center}
\begin{tabular}{lccccc}
\hline\hline
Region & $\bar{U}$ & Average $U$ for & $U\!V\!F$ - $\bar{U}$  & $U\!V\!F$ & $U\!V\!F$ \\
             &                & vortex formation &                                     & (non-         & (shed) \\
             &                & $U\!V\!F$            &                                     &  shed)       &            \\

\hline
SN  & 1.5  & -1.3 & -2.8  &-0.8 &  -1.7\\
\hline
MP & 0.0  & -2.7 & -2.7 &-2.2 & -3.7\\
\hline
SS & -0.3  & -3.1 &  -2.8 & -3.1& -2.9\\
\hline
JV & -1.1 & -3.8 &  -2.7 & -2.3& -4.8\\
\hline
\hline
\end{tabular}
\end{center}
\end{table}

The annual cycle of monthly vortex count for vortices originating from terrain-induced circulations in regions SN and MP are shown in Fig.\ 4. As seen here, vortex counts maximize in
the months with the strongest mean easterlies with a peak in January over SN and in February in MP. During the boreal summer months, prevailing westerlies over these regions result in the production of few lee vortices. In contrast over the SS and JV regions, the vortex count peaks in the boreal summer months when easterlies prevail over these southern regions (Fig.\ 5). Also,
while the northern regions have only a single peak in production numbers, the annual cycles over the southern regions appear more complex with multiple peaks. For example, over SS there is a gradual ramp up to a peak in August when easterlies ($U_S$) are strongest, but then the highest peak occurs in October when mean easterlies are reduced from the late summer maximum. Likewise, over JV multiple peaks occur with the highest number in July and a secondary peak in November. And while the mean winds have transitioned to weak westerlies by November, the mean low-level vorticity remains negative (i.e., cyclonic). Also, listed in Figs.\ 4 and 5 are the correlation between the monthly-averaged vortex frequency and monthly-averaged zonal winds. Correlations are near or above $-0.9$ in all the regions except SS indicating that the annual cycle of zonal winds in the former regions explain $\sim$80\% of the variance in monthly vortex counts. On the hand, the lower correlation in region SS ($-0.67$) suggests that zonal wind variability outside the annual cycle (i.e., intraseasonal or interannual time scales) is more influential on the monthly vortex count in this region.

\begin{figure}[thbp]
\centerline{\includegraphics[width=3in]{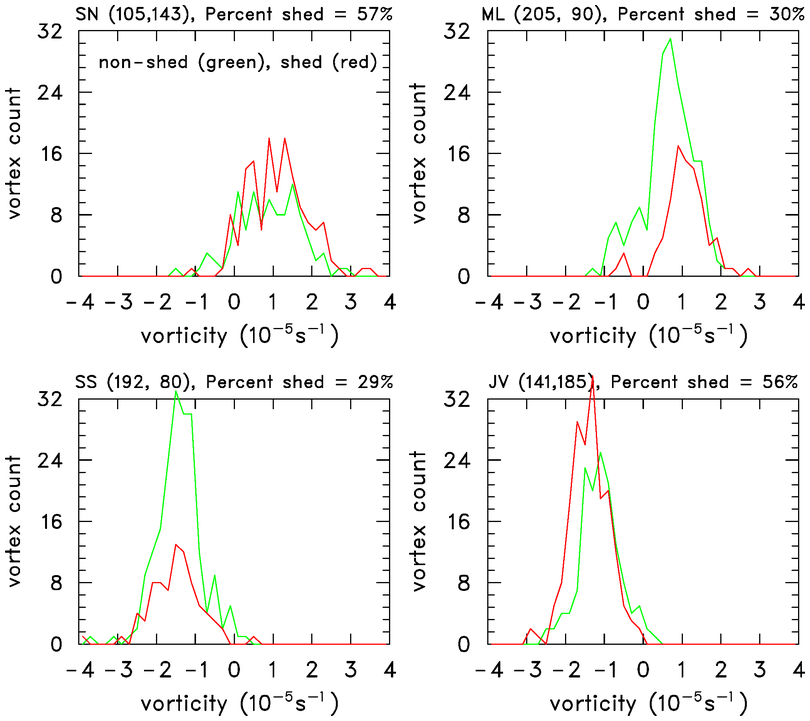}}

\protect\caption{
Vortex count as a function of relative vorticity (10$^{-5}$ s$^{-1}$) for non-shed (green) and shed vortex (red) cases for the four regions of interest in this study. Numbers in parentheses in top labels represent the number of non-shed and shed vortex cases, respectively.
}\label{f7}
\end{figure}

\begin{table}[t]

\caption{
Relative vorticity (10$^{-5}$ s$^{-1}$) statistics for regions of interest. From left to right: Mean zonal vorticity for region ($\bar{\zeta}$) based on 2008-17 period, average $\zeta$ associated with vortex formation ($V\!V\!F$), difference between $V\!V\!F$ and $\bar{\zeta}$, $V\!V\!F$ for non-shed vortices, $V\!V\!F$ for shed vortices.
}
\label{t5}
\begin{center}
\begin{tabular}{lccccc}
\hline\hline
Region & $\bar{\zeta}$ & Average $\zeta$ for & $V\!V\!F$ - $\bar{\zeta}$  & $V\!V\!F$ & $V\!V\!F$ \\
             &                & vortex formation &                                     & (non-         & (shed) \\
             &                & $V\!V\!F$            &                                     &  shed)       &            \\

\hline
SN  & -0.07  & 1.2 & 1.27  & 1.1 &  1.3\\
\hline
MP & -0.05  & 1.0 & 1.05 & 0.8 & 1.3\\
\hline
SS & -0.68  & -1.2 &  -0.52 & -1.2& -1.3\\
\hline
JV & -0.56 & -1.1 &  -0.54 & -0.9& -1.2\\
\hline
\hline
\end{tabular}
\end{center}
\end{table}

To further explore the relationship of vortex formation and shedding to winds and vorticity, Figs.\! 6 and 7 examine the distribution of the zonal winds and vorticity for shed and non-shed cases over the four regions of interest, respectively. In these figures the wind speed and vorticity represent their values at the time the vortex was first identified in the regions of interest. Included in Fig.\! 6 are percentage of non-shed and shed vortices occurring with easterly flow. In addition, Tables 4 and 5 summarize the zonal wind and vorticity statistics for vortex formation in these regions. Clearly the presence of easterly flow is a dominant factor in determining vortex
formation and shedding. JV, which had the highest number of vortices during this 10-yr period (326), also had the strongest easterlies during times of vortex formation. In contrast, SN had the
fewest vortices (248) and the weakest easterlies. The increase in the easterly component of the wind from its mean state and that observed with vortex formation is remarkably consistent among the regions being 2.75 m s$^{-1} \pm$ 0.05 m s$^{-1}$ (Table 4, $U\!V\!F - \bar{U}$). This consistency is likely coincidental, as the analogous cyclonic vorticity increase (Table 5, $V\!V\!F - \bar{\zeta}$) varies considerably among the regions being nearly double for the NH locations. In general, shed cases are associated with stronger easterlies and higher cyclonic vorticity (Tables 4 and 5). While non-shed cases occur with easterly flow 61-81\% of the time, the presence of easterly flow for shed cases, as one might expect, is even higher (76\% to 94\%). As seen in Table 1 the shed frequency, i.e., percentage of vortices which shed and move westerly after forming, varies among the regions ranging from $\sim$30\% in ML and SS to greater than 50\% in SN and JV. While SN has the highest shed frequency at 57\%, it also has the weakest easterly winds, such that factors the other than zonal wind speed (e.g., vertical shear, large-scale thermodynamical environment) must be contributing to the frequency at which vortices stay intact and shed once they form.

\subsection{Relationship of vortex production to large-scale signals (MJO, IOD, ENSO)}

The 10-yr time series of monthly vortex counts for the four regions of interest are shown in Fig.\! 8. Beyond the seasonal variability, as depicted in the Figs.\! 4 and 5, these time series contain fluctuations on intraseasonal and interannual time scales as well. Leveraging this longer analysis period, we now consider variability of vortex formation and its relationship to various large-scale signals such as the MJO, IOD, and ENSO signals.

\begin{figure}[thbp]
\centerline{\includegraphics[width=3in]{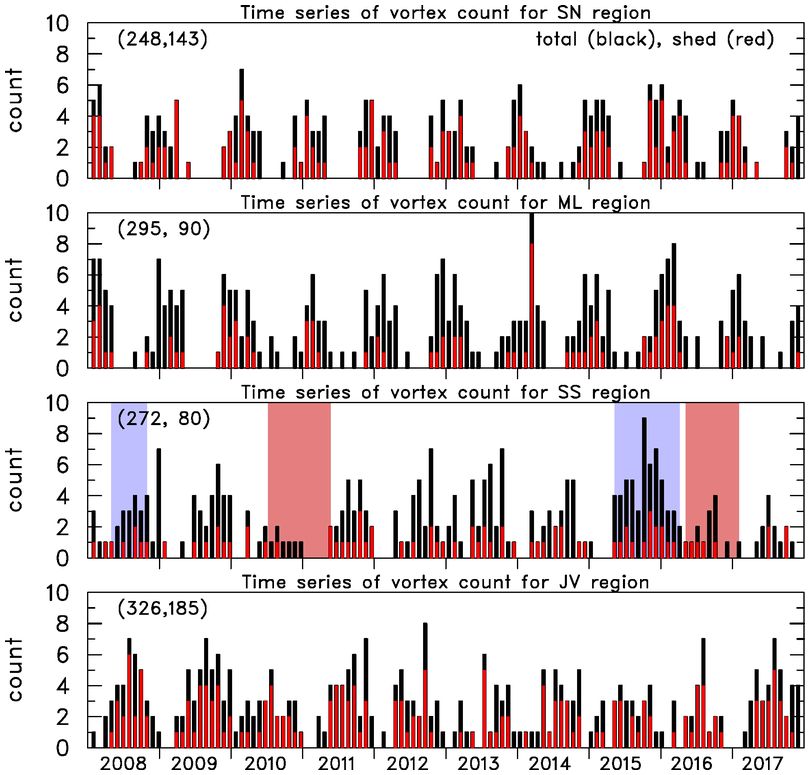}}

\protect\caption{
Time series of monthly vortex count for four regions of interest (black bars show total count, red bars show shed-vortex count). Numbers in parentheses give the total and shed vortex count for 2008-17 period. Blue (red) shading represent periods in region SS that experienced prolonged periods of anomalous negative (positive) monthly zonal wind anomaly $U^\prime$.
}\label{f8}
\end{figure}

Figure 9 examines the vortex and zonal wind statistics based on the phase of the MJO. Here vortex counts and winds are only considered if the MJO signal was reasonably strong, defined here as the RMM index amplitude being $\ge 1$. In general, vortex counts are highest when easterlies are strongest (i.e. phases 7, 8, and 1). The modulation of vortex count by MJO phase is most prominent in the southern regions (SS and JV) where the zonal wind speed variations by the MJO are more pronounced. For example, the wind speed change over the course of the MJO over the southern regions is about 4-5 m s$^{-1}$ compared to 2-3 m s$^{-1}$ over SN and MP. Consistent with the findings above, the right panels of Fig.\! 9 show that shed vortices are generally associated with stronger easterly winds.

\begin{figure}[thbp]
\centerline{\includegraphics[width=3in]{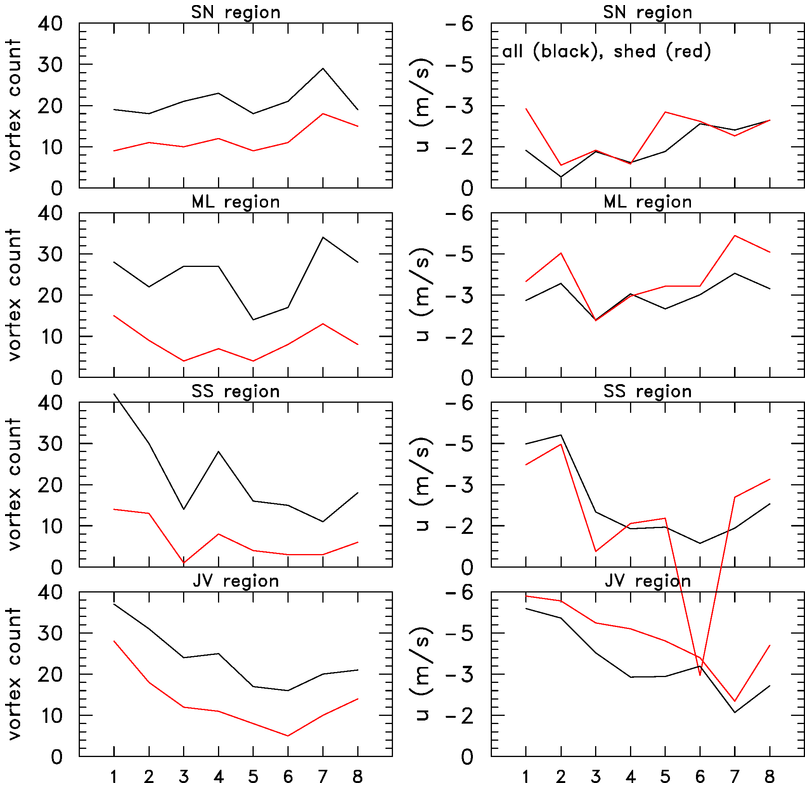}}

\protect\caption{
Statistics for vortex count (left panels) and winds (right panels) for four regions of interest (black curves are for all vortices, red curves are for shed vortices) as function of MJO phase.
}\label{f9}
\end{figure}

A common feature associated with the MJO over the IO is a region of strong near-equatorial westerlies which follows the period of strongest convection. This feature, referred to as a westerly wind burst (WWB) and examined extensively in previous literature \citep{rui1990development, hendon1994life, kiladis1994large} has a prominent low-level signature peaking near 700 hPa  \citep{KiladisEA2005}. Since F16 examined a case where vortex formation in both hemispheres appeared to be assisted by blocking of strong low-level equatorial westerlies by central Sumatra, Fig.\! 10 considers the statistics of vortex formation and low-level equatorial winds based on the MJO phase. Here the statistics for all the regions are considered together and the low-level equatorial flow $U_E$ is taken from the region (2$^{\circ}$N-2$^{\circ}$S, 97$^{\circ}$E) shown in Fig.\! 1, that is centered on the equator and a few hundred kms to the west of Sumatra. As seen here the vortex counts are highest during MJO phase 1 when equatorial westerlies are weak and off equatorial winds are easterly (as seen in Fig.\! 9). However, a secondary peak occurs in vortex count during phase 4 when the equatorial westerlies are strongest. This peak in westerlies over $U_E$ during MJO phase 4 is consistent with the canonical structure of the MJO described in Kiladis et al.\! (2005) which places the low-level westerlies during this phase to the west the MJO convective center located over the Maritime continent. This suggests that westerly surges on the equator associated with MJO convection may be contributing to the uptick in vortex formation during MJO phase 4. This uptick during MJO phase 4 is seen in all regions but is most prominent in SS and is weakest in JV (Fig.\! 9). The mechanism for vortex formation in this situation is most likely a combination or superposition of two effects: (1) the generation of equatorial Rossby waves to the west of the MJO convective envelope \citep{Gill1980} and (2) blocking and splitting of the flow as it encounters the steep terrain of Sumatra, which can serve to enhance counter-rotating circulations at opposite ends of the island \citep{smith1989mountain}.

\begin{figure}[thbp]
\centerline{\includegraphics[width=3in]{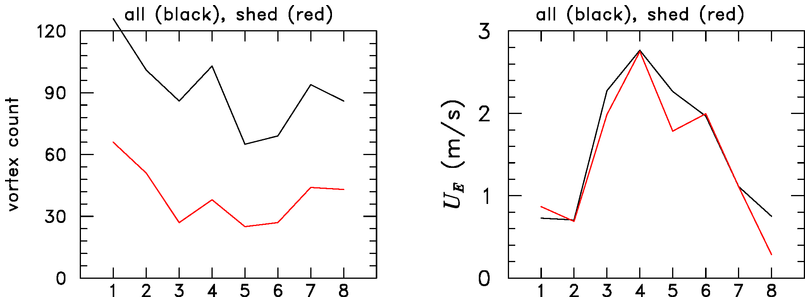}}

\protect\caption{
Statistics for vortex count (left panels) and $U_E$ winds (right panels) for all regions collectively (black curves are for all vortices, red curves are for shed vortices) as function of MJO phase.
}\label{f10}
\end{figure}

\begin{figure}[thbp]
\centerline{\includegraphics[width=3in]{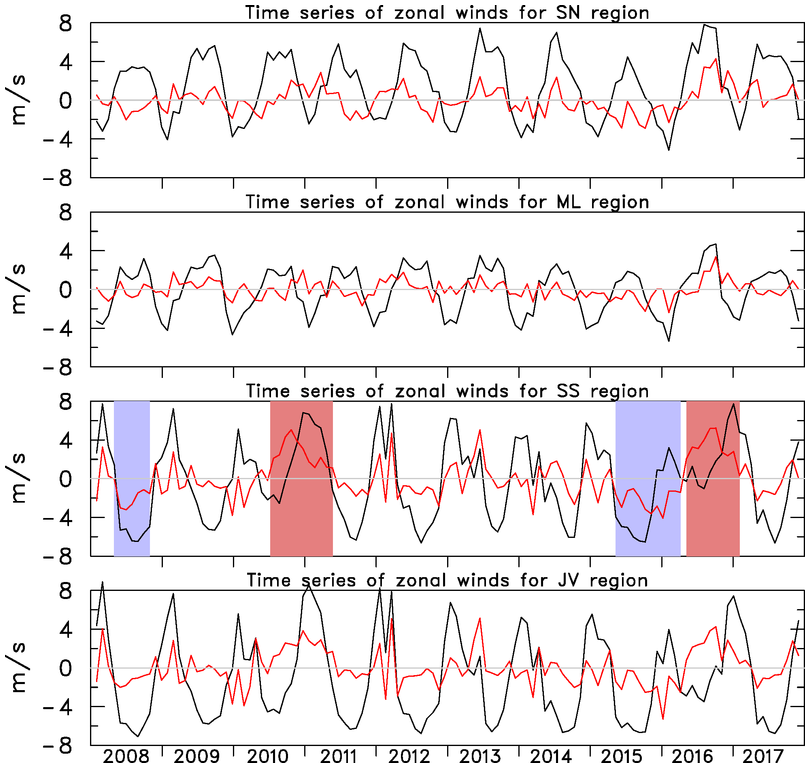}}

\protect\caption{
Time series for period 2008-17 of monthly-averaged zonal wind (black curve) and monthly anomaly of zonal wind from its annual cycle ($U^\prime$, red curve) for regions of interest from top to bottom: SN, MP, SS, and JV. Blue (red) shading represent periods in region SS that experienced prolonged periods of anomalous negative (positive) $U^\prime$.
}\label{f11}
\end{figure}

To further examine the temporal variability in the vortex count depicted in Fig.\! 8, we show in Fig.\! 11 the time series of monthly-averaged zonal wind and its anomaly from the annual cycle (depicted in Figs.\! 4 and 5) for the four regions of interest. The large, negative correlations of these monthly-averaged winds to vortex count (Table 6, middle column) reflect the strong relationship that exists between vortex formation and the presence of easterly winds. The variance explained by this relationship is smallest in region SS consistent with its behavior over the annual cycle discussed earlier. However, when vortex count is regressed against the anomaly of the zonal winds from its annual cycle, this relationship is strongest in region SS (Table 6, rightmost column). This suggests that perturbations to the annual cycle of winds caused by signals other the annual cycle are influencing vortex formation in region SS. For whatever reason, periods of persistent anomalous winds are more prevalent in the southern regions and largest in SS. Four such periods (two with negative wind anomalies and two with positive anomalies) have been highlighted in Figs. 8 and 11 to show how anomalous zonal winds are impacting vortex formation in region SS. For these two periods with anomalous easterlies (-2.1 m s$^{-1}$) the SS vortex count averaged 4/month (76 vortices in 19 months); for the two periods with anomalous westerlies (+2.6 m s$^{-1}$), the SS vortex count averaged 1.1/month (23 vortices in 22 months). This fourfold increase in vortex formation between anomalous wind periods highlights the sensitivity of vortex development to anomalies in the large-scale wind field.

\begin{table}[t]

\caption{
(middle) Correlation (r) between monthly vortex count and monthly-averaged zonal wind ($\bar{U}$ ), and (right) correlation between monthly vortex count and monthly-zonal wind anomaly from its mean annual cycle ($U^\prime$). For these correlations winds averaged over each region of interest were used based on data for 2008-2017 period.
}
\label{t6}
\begin{center}
\begin{tabular}{lcc}
\hline\hline
Region & r (vortex count, $\bar{U}$) & r (vortex count, $U^\prime$)   \\
\hline
SN & -0.77 & -0.15  \\
\hline
MP & -0.79 & -0.20 \\
\hline
SS & -0.59 & -0.41 \\
\hline
JV & -0.70 & -0.27 \\
\hline
\hline
\end{tabular}
\end{center}
\end{table}

Having established the strong dependence of vortex formation on the large-scale wind field, we now consider how the wind field is impacted by circulation changes related to the IOD and ENSO signals. Figure 12 shows the time series of the ENSO MEI.v2 and IOD indices for the 10-yr period of this study. These two indices show a weak temporal correlation (0.23) during this period. Detailed discussions of their interaction and influence on each other can be found in other studies \citep{Behera2006cgcm, Cai2011interactions}. The 2008-17 time frame contained two prominent ENSO warm events near the end of 2009 and 2015 into early 2016 and two cool events in 2008 and the latter half of 2010 into 2011. In general, the warm (cool) events are associated with suppressed (enhanced) convection over the Maritime Continent and enhanced low-level easterlies (westerlies) over the IO \citep{ramage1968role, chang2004relationship}. The negative phase of the IOD as depicted in the schematic in Fig.\! 13 is associated with warm SSTs in the eastern IO, increased convection over the maritime continent, and enhanced low-level westerlies over the equatorial IO. The positive phase of the IOD would have opposite impacts, namely suppressed convection over the maritime continent and enhanced low-level easterlies over the equatorial IO.

\begin{figure}[thbp]
\centerline{\includegraphics[width=3in]{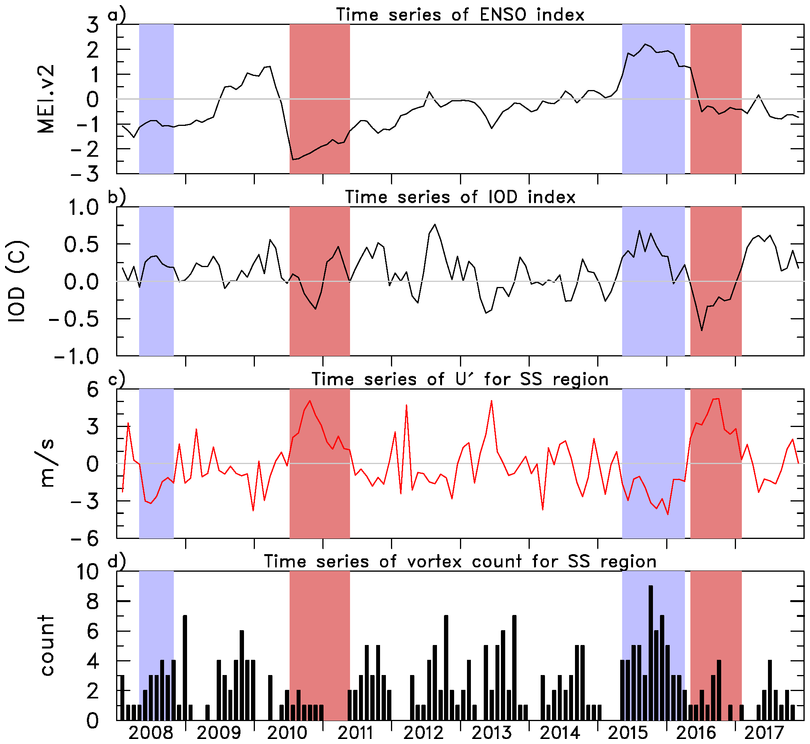}}
\protect\caption{
Times series for period 2008-17 of: (a) ENSO index, (b) IOD index, (c) $U^\prime$ over SS, (d) vortex count over SS. Blue (red) shading represent periods in region SS that experienced prolonged periods of anomalous negative (positive) $U^\prime$.
}\label{f12}
\end{figure}

\begin{figure}[thbp]
\centerline{\includegraphics[width=3in]{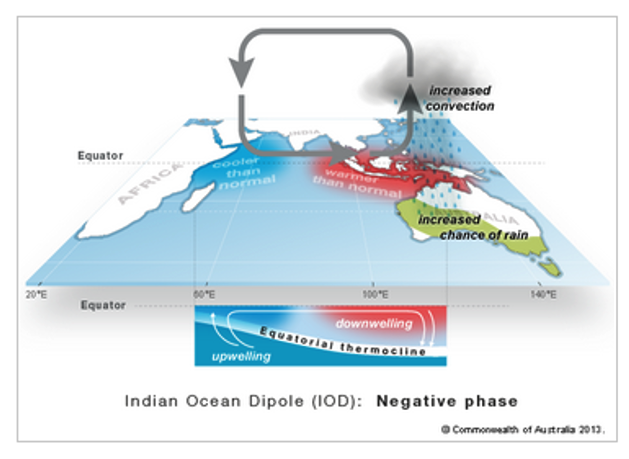}}
\protect\caption{
Schematic of positive phase of Indian Ocean Dipole (from NOAA Climate.gov).
}\label{f13}
\end{figure}

Table 7 lists the correlations between monthly zonal wind anomaly ($U^\prime$) and the ENSO and IOD indices. In all the regions a small negative correlation is observed, which suggests that positive (negative) ENSO and IOD conditions are driving anomalous easterlies (westerlies). Correlations are slightly larger between $U^\prime$ and the IOD index in all regions except JV, possibly because the IOD is a more regional index for the areas considered here. Having identified the importance of the wind field on vortex formation and the relationship, albeit weak, between these large-scale indices and the winds, we now consider whether vortex formation is related to the large-scale circulation changes associated with ENSO and the IOD signals. The right-most two columns in Table 7 list the correlations between vortex count and these large-scale indices. Only in region SS are correlations significant, which is consistent with the stronger relationship between $U^\prime$  and vortex count seen in Table 6 and the higher correlations between $U^\prime$ and indices in Table 7 for this region. The positive correlations seen in region SS suggest that positive (negative) ENSO and IOD conditions are associated with enhanced easterly (westerlies) and thus a higher (lower) rate of vortex formation. The bottom two panels of Fig.\! 12 show the $U^\prime$ and vortex count for region SS to help illustrate the relationship between these fields. It is worth noting that when the ENSO and IOD indices are of the same sign as in the 2015-16 period, the anomalous winds and impact on the vortex formation seem most dramatic. Vortex formation in region SS typically peaks in the months from May to October (Fig.\ 5) when the mean flow has an easterly component. For these months, Fig.\! 14a depicts the total flow field during 2015 when enhanced easterlies were present over SS and 30 vortices formed. In contrast, anomalous westerlies during the 2016 period (Fig.\! 14b) resulted in weak flow from the northwest over SS and the formation of only 12 vortices. The reason that this relationship is significant only in region SS during this period is unclear at this time.

\begin{table}[t]

\caption{
(from left to right) Correlation (r) between monthly-zonal wind anomaly ($U^\prime$) and ENSO index, correlation between $U^\prime$ and IOD index, correlation between number of vortices formed and ENSO index, and correlation between number of vortices formed and IOD. All correlation were computed using data for the 2008-2017 period.
}
\label{t7}
\begin{center}
\begin{tabular}{lcccc}
\hline\hline
Region & r ($U^\prime$, ENSO)  & r ($U^\prime$, IOD)  & r(\# vor, & r(\# vor,  \\
            &                                      &                                   &  ENSO) &      IOD) \\
\hline
SN & -0.35  & -0.41 & 0.10 & 0.13 \\
\hline
MP & -0.25 & -0.33 & 0.11 & <0.10 \\
\hline
SS & -0.45 & -0.50 & 0.36 & 0.21 \\
\hline
JV & -0.46 & -0.35 & <0.10 & <0.10 \\
\hline
\hline
\end{tabular}
\end{center}
\end{table}

\begin{figure}[thbp]
\centerline{\includegraphics[width=3in]{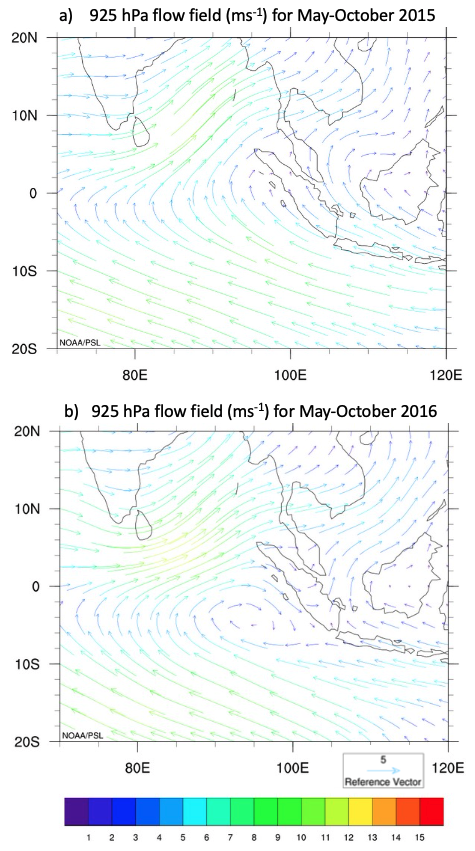}}
\protect\caption{a) 925 hPa flow field for May – October 2015, b) 925 hPa flow field for May – October 2016. Reference vector and scale for wind vectors (m s$^{-1}$) shown at bottom. Red box outlines region SS. Created using NCEP reanalysis II and NOAA/PSL visualization tools.
}\label{f14}
\end{figure}

\subsection{Characteristic of the long-term TC record over the IO}

One of the key findings of F16 is that 25\% of the TCs in the IO basin during the YD period had their origins in vortices that shed from topographic features in the western maritime continent. A cursory examination of remaining 7.5 years in the 2008-17 period found an additional 15 TCs that had their origins in shed vortices. Most likely this number is an undercount since vortices can weaken below the vorticity tracking threshold (1$\times$10$^{-5}$s$^{-1}$) before reintensifying. While identification of these more subtle cases is beyond the scope of this study, it is instructive to examine some of the characteristics of the long-term TC record over the IO.

\begin{figure}[thbp]
\centerline{\includegraphics[width=3in]{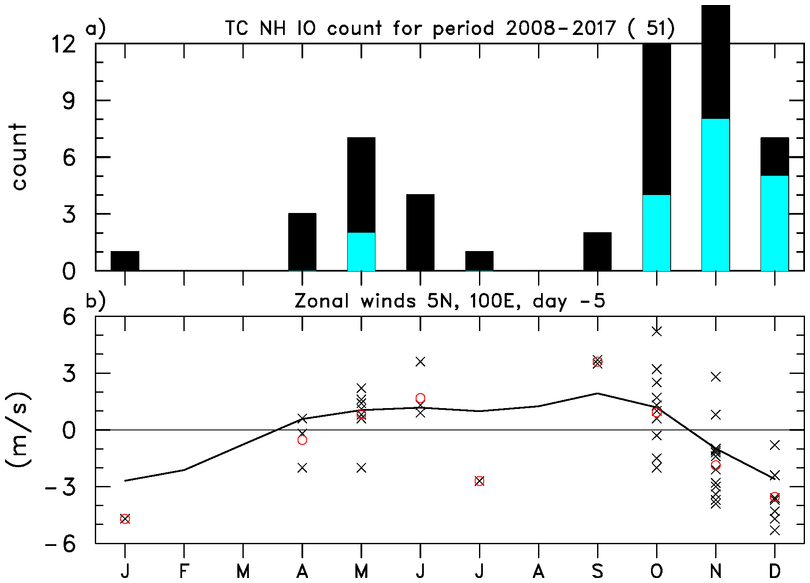}}
\protect\caption{
 (a) Monthly frequency of tropical cyclones over the NH IO (black) and impacting India and Sri Lanka (cyan) for the period 2008-2017. (b) Monthly mean zonal wind (m s$^{-1}$) over northern Sumatra ($U_N$, black curve), zonal wind speed $U_N$ at 5 days prior to a storm initialization (one x-symbol for each TC), the red circle shows the mean of winds represented by the x- symbols in that month.
}\label{f15}
\end{figure}

\begin{figure}[thbp]
\centerline{\includegraphics[width=3in]{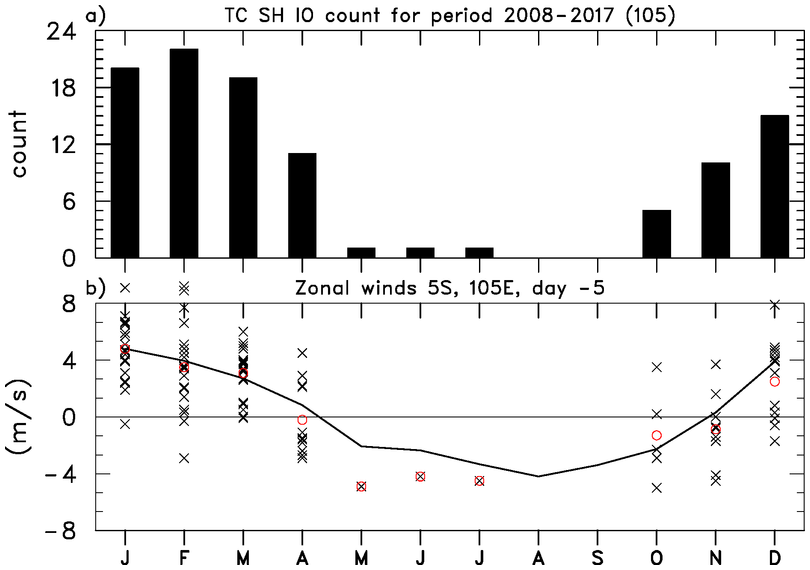}}
\protect\caption{
 (a) Monthly frequency of tropical cyclones over the SH IO (black)  for the period 2008-2017. (b) Monthly mean zonal wind (m s$^{-1}$) over northern Sumatra ($U_S$, black curve), zonal wind speed $U_S$ at 5 days prior to a storm initialization (one x-symbol for each TC), the red circle shows the mean of winds represented by the x- symbols in that month.
}\label{f16}
\end{figure}

Figs.\! 15 and 16 show the annual cycle of TC count over the IO for the NH and SH regions for 2008-17, respectively. During this 10-yr period, 51 TCs were identified over the NH and 105 over the SH. Of the NH cases, 22 made landfall in India or Sri Lanka with TC intensity or as decaying TCs. Over the NH IO, TC numbers peak in November with a secondary maximum in May. While the NH May peak occurs during a period of mean westerlies over ($U_N$), the November maximum occurs during a period of climatological easterlies (Fig.\! 15b). Also plotted in this figure are the winds over the northern region ($U_N$) five days prior\footnote{Winds over the regions of interest ($U_N$ and $U_S$) at 5 days prior to time $t_0$ were used here because over the period from $t_0$ to $t_0 -$15 days they had the strongest easterly component.} to the time ($t_0$) when the cyclone was first identified by JTWC. It worth noting that while in some months the $U_N$ winds are mean westerly, there are a number of cases in which the winds were easterly at $t_0$ such that the potential for the TC to have initialized from a lee vortex existed. Over the SH IO, TC numbers peak in February during which $U_S$ is mean westerly (Fig.\! 16). During the months with mean easterlies in $U_S$ (May through October) the TC count is generally low. Similar to the NH, there are number of TC cases in months with mean westerlies in which $U_S$ winds were easterlies at $t_0$.

\begin{figure}[thbp]
\centerline{\includegraphics[width=3in]{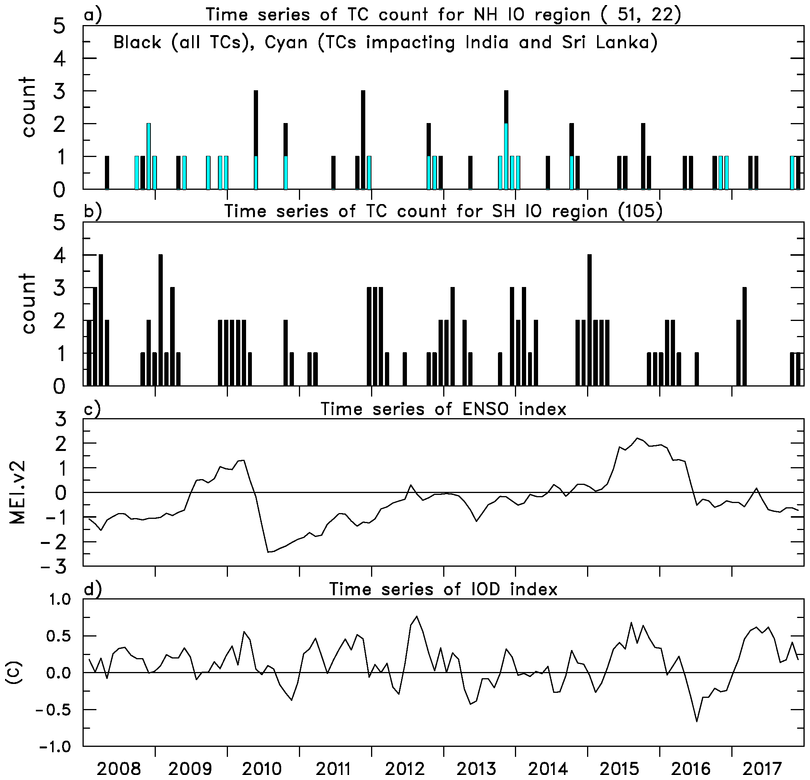}}
\protect\caption{
Time series for period 2008-17 of: a) monthly count for NH IO TCs (black), TCs impacting India and Sri-Lanka (cyan), b) monthly count for SH IO TCs (black), c) ENSO index, d) IOD index. Numbers in parentheses indicate the number of TCs in each category for this 10-yr period.
}\label{f17}
\end{figure}

The time series of monthly TC count over the NH and SH IO regions is shown in Fig.\! 17, along with the ENSO and IOD indices for the 2007-2008 period. No significant correlations (<0.1) exist between the TC count time series and these indices. However, if we consider only TCs that originated east of 90$^\circ$E, a small negative correlation ($-0.2$) is found between NH TCs and the IOD. This suggests that the warmer waters over the eastern IO (negative phase of IOD) contribute to higher TC numbers.

Since other ocean basins have shown a sensitivity of TC formation to the phase of the MJO \citep{maloney2000modulation, zhang2013madden}, we now composite the IO TC count by the MJO phase based on their initialization time $t_0$. We note that time from $t_0$ to TC formation is on average 36 hours over the NH and 45 hours over the SH. Using these criteria, Fig.\! 18 shows that in both hemispheres cyclone initiation occurs most frequently in MJO phase 3. These findings are consistent with \cite{bessafi2006modulation} who found that TC numbers over the southern IO increase by a factor of 2.6 between MJO convectively active phases (2 and 3) compared to convectively suppressed phases (8 and 1). Recalling that easterlies and vortex formation peak around phase 1 (Fig.\! 10)  or about 8 to 12 days prior to MJO phase 3, and that the average gestation period from vortex formation to TC genesis is 10.1 days (F16), then shed vortices conceivably are contributing to the storm initialization peak in MJO phase 3. For TCs impacting India and Sri Lanka (Fig.\! 18b), the storm initialization peak occurs slightly later around MJO phase 4 in agreement with the findings of \cite{hall2001modulation}. This slightly later peak over the northern IO is due to the poleward drift with time of the convectively active regions of the intraseasonal signal in this region of the world.

\begin{figure}[thbp]
\centerline{\includegraphics[width=3in]{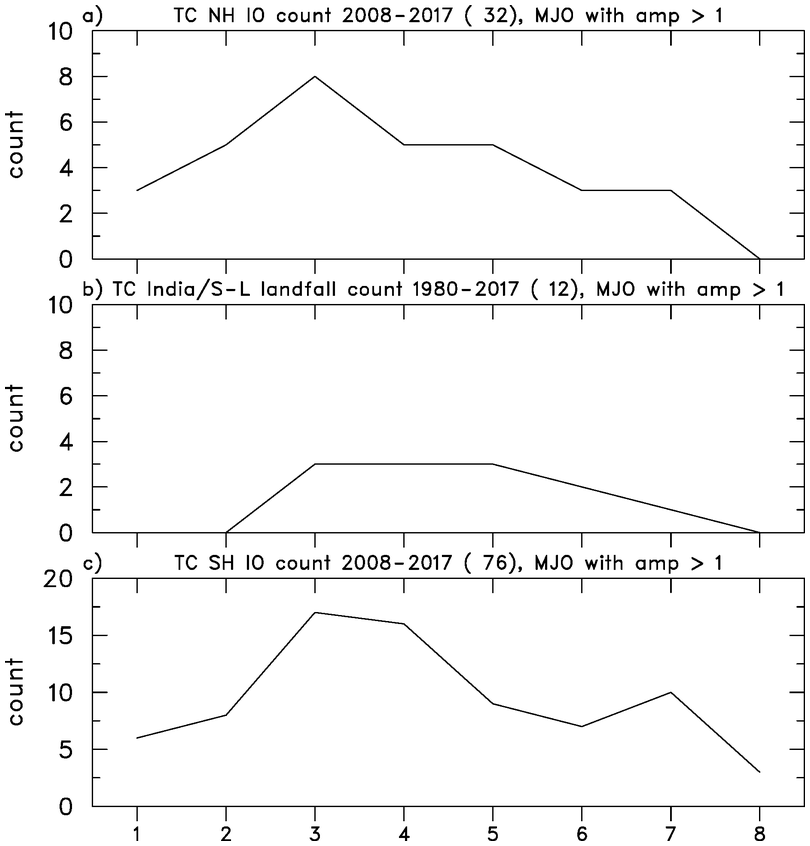}}
\protect\caption{
Statistics for TC count as function MJO phase based on period 2008-17 for (top) NH IO TCs, (middle) TCs impacting India and Sri-Lanka and (bottom) SH IO TCs. Time that storm was first identified by JTWC was used to categorize TC by MJO phase. Only cases where the MJO amplitude was > 1 are considered here.
}\label{f18}
\end{figure}

\begin{figure}[thbp]
\centerline{\includegraphics[width=3in]{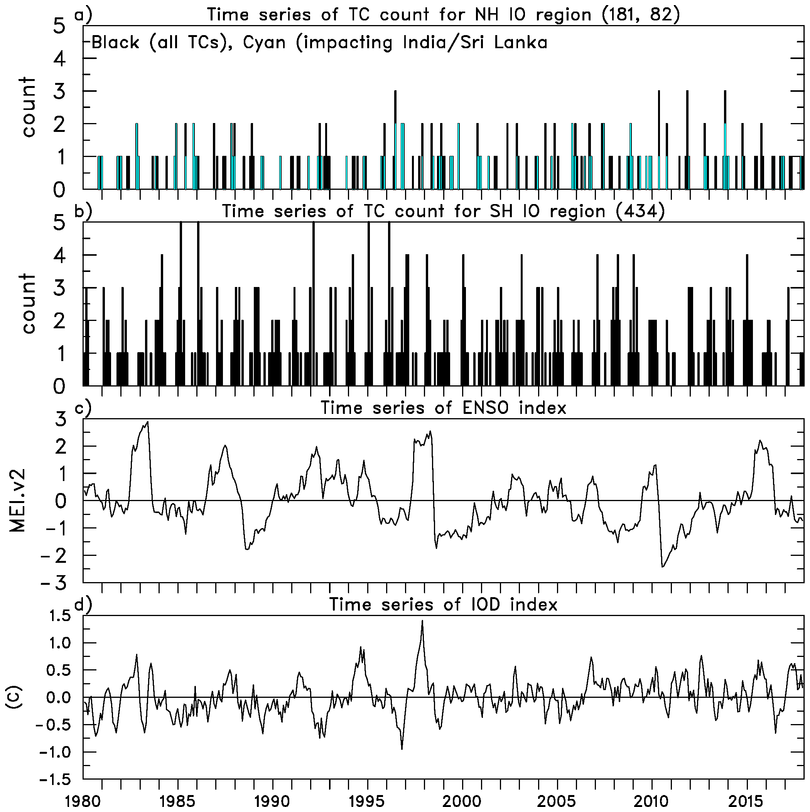}}
\protect\caption{
Time series for period 1980-2017 of: a) monthly count for NH IO TCs (black), TCs impacting India and Sri-Lanka (cyan), b) monthly count for SH IO TCs (black), c) ENSO index, d) IOD index. Numbers in parentheses indicate the number of TCs in each category for this 38-yr period.
}\label{f19}
\end{figure}

\begin{figure}[thbp]
\centerline{\includegraphics[width=3in]{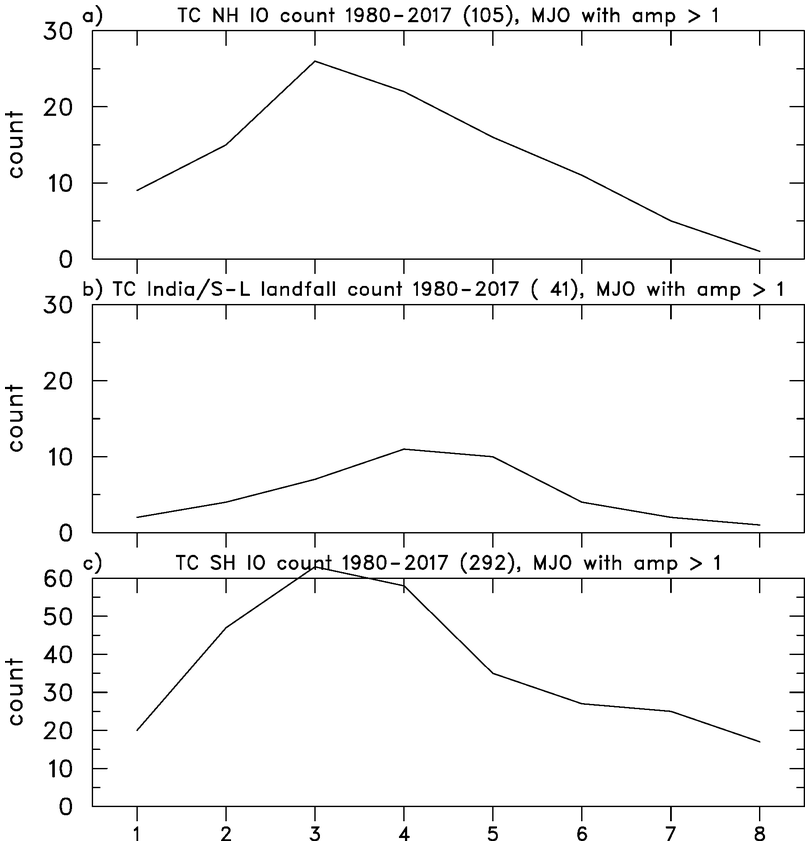}}
\protect\caption{
Statistics for TC count as function MJO phase based on period 1980-2017 for (top) NH IO TCs, (middle) TCs impacting India and Sri-Lanka and (bottom) SH IO TCs. Time that storm was first identified by JTWC was used to categorize TC by MJO phase. Only cases where the MJO amplitude was > 1 are considered here.
}\label{f20}
\end{figure}

To examine the representativeness of the 10-yr TC record relative to a longer period, Fig.\! 19 shows time series of TC count for years 1980-2017. This period contained several ENSO cycles with significant El Nino (warm) events in 1982-83, 1987-88, 1997-98, and 2015-16. During this 38-yr period the NH IO basin averaged 4.8 TCs/year while the SH IO basin experienced more than double this rate at 11.4 TCs/year. These rates are similar to the 2008-17 period but below that of the YD period investigated by F16 (Table 6). Landfalling or decaying TCs impacting India and Sri Lanka averaged 2.2 per year during this period. No long-term trends were detectable in any of these time series. Also, no significant correlation (>0.1) of TC numbers to large-scale indices were observed except for a slight correlation ($-0.2$) between TC numbers for storms initializing east of 90$^\circ$E and the IOD index. As in the 10-yr record (2008-17), TC initializations for this longer period (Fig.\! 20) peak in MJO phase 3 over both NH and SH IO basins and MJO phase 4 for storms impacting India/Sri Lanka.

\section{Summary and conclusions}

With the recent availability of the ERA5 high-resolution reanalyses for the years 1979 to present, the opportunity to extend the Fine et al.\! (2016, F16) vortex tracking analyses to additional years became available. The goal of this study is to determine the representativeness of F16’s findings compared to a longer period (2008-2017) using this improved operational analysis and to examine the characteristics of vortex formation and shedding and their relationship to intraseasonal and interannual oscillations for this longer record.

F16 identified four vortex genesis regions near the northern and southern tips of Sumatra (see Fig.\! 2 for their locations). Application of the tracking software to 10 years of ERA5 analyses in a similar fashion to F16 yielded the following conclusions based on statistics for these four vortex “hotspots”.

\begin{itemize}

\item
For the 2.5-yr period, referred herein as the YD-period since F16 used YOTC/DYNAMO (YD) analyses, terrain-induced production of vortices was similar between F16 and ERA5 (123/yr) while the vortex shedding rate was slightly higher in ERA5 (42\% versus 33\% in F16).

\item
For the 10-yr period, vortex production in ERA5 was 114/yr; vortex shedding frequency was 44\%.

\item
The higher rate (8\%) of vortex formation during the YD-period compared to the longer-term mean was likely due to an enhanced easterly component of the flow during the YD-period.
\end{itemize}

While the YD-period of F16’s analyses was slightly more active in terms of vortex production than the 10-yr period, in general F16’s findings are representative of the longer-term record. The following represents conclusions based on additional analysis of the 10-yr record of vortex production and shedding in the ERA5 dataset:

\begin{itemize}

\item
$\sim$70\% ($\sim$80\%) of all vortices (shed vortices) occur with easterly low-level (925 hPa) flow over the regions of interest.
\item
Vortices which shed (i.e., move westward over the IO) are typically associated with stronger low-level easterlies and higher cyclonic vorticity. However, it is unclear why the SN region (the area off the northern tip of Sumatra) has the highest shed frequency yet is characterized with the weakest easterlies.
\item
Vortex counts are highest near MJO phase 1 when stronger low-level easterlies are present over the maritime continent region.
 \item
The annual cycle of zonal winds largely controls vortex formation rates in the regions of interest. For reasons unclear at this time, this relationship is less apparent for region SS (the area off the southern tip of Sumatra) where zonal wind variability outside the annual cycle plays a more impactful role in determining vortex formation.
\item
Low-level equatorial westerlies west of Sumatra, which peak during MJO phase 4, are coincident with a secondary peak in vortex (and shed vortex) counts. This suggest that low-level westerly surges on the equator impinging on Sumatra associated with the MJO contribute to an increase in wake vortex development.
 \item
Periods with anomalous low-level easterlies (westerlies) have higher (lower) than average vortex counts.
\item
For reasons unclear at this time, anomalous low-level winds were most pronounced in region SS. For example, during periods of anomalous easterlies (19 months), the vortex count averaged 4/month. In contrast during periods of anomalous westerlies (22 months), the vortex count averaged 1.1/month. This fourfold increase in vortex formation between periods of anomalous low-level zonal winds highlights the sensitivity of vortex development to anomalies in the large-scale wind field.
\item
While positive (negative) El-Nino-Southern Oscillation (ENSO) and Indian-Ocean Dipole (IOD) conditions drive anomalous easterlies (westerlies), the relationship between vortex count and these large-scale indices is generally weak (r < 0.1). Only in region SS do the large-scale conditions reflected in these indices appear to impact vortex formation numbers (r=0.36 with ENSO and 0.21 with IOD). These positive correlations imply that the positive ENSO and IOD conditions drive anomalous easterlies which result in higher vortex formation rates.
\end{itemize}

While identifying which Indian Ocean (IO) tropical cyclones (TCs) were associated with low-level topographically forced shed vortices was beyond the scope of this study, the following TC statistics were examined:
\begin{itemize}
\item
Based on the Joint Typhoon Warning Center (JTWC) database, 51 (105) TCs were identified
over NH (SH) IO during 2008-17 period. This TC frequency (15.6/yr) was somewhat less than that observed during the F16 period (19.6/yr). Determining the percentage of TCs in the long-term record which originated from shed vortices is left for a future study.
\item
TC initialization is most frequent over the IO during MJO phase 3 when convective anomalies are largest over the eastern IO. Since shed vortex counts peak around phase 1 (Fig.\ 10) or about 8-12 days earlier, it is possible that shed vortices contribute to the TC initialization peak observed in MJO phase 3.
\item
Little, if any, correlation exists between TC frequency in the IO and ENSO and IOD signals. A slight uptick in TCs numbers for storms originating east of 90°E during the negative IOD phase suggest that warm waters in eastern IO may be contributing to this increase.
\end{itemize}

Future work in the area of this study could include extending the tracking analyses to yet additional years to determine whether the relationships summarized above are robust over yet longer periods. Remaining questions from this study, which further analysis could help elucidate, include the following. What factors, other than zonal wind, control the frequency of vortex shedding? Why do the large-scale signals (such as ENSO and IOD) have a preferred influence on the zonal winds in region SS resulting in more prominent zonal wind anomalies and a larger impact on vortex formation in this region? What percentage of shed vortices in the extended period (2008-2017) developed into TCs, and in particular, how many of these TCs impacted vulnerable populations in India and Sri Lanka? Addressing this latter question may require that the vortex tracking code be run with geopotential height anomalies and/or a smaller vorticity threshold to track shed vortices which weaken then subsequently reintensify. And finally, Kuettner’s (1989) observations of IO twin cyclones could be examined to determine how many of these cases can be attributed to twin shed vortices \citep{ferreira1996dynamical}.

 \acknowledgments
We thank Kevin Hodges from the University of Reading for his assistance in getting the tracking code working with ERA5 analyses and Rick Taft for his IT support with this project. This research has been supported by the National Science Foundation under Grant AGS-1853633.

\datastatement
The ERA5 data were obtained from: https://cds.climate.copernicus.eu, the MJO RMM indices were provided by the Australian Bureau of Meteorology from: www.bom.gov.au/climate/mjo, the MEI.v2 ENSO index from: https://psl.noaa.gov/enso/mei/, the DMI data from: https://psl.noaa.gov/gcos\underline{~~}wgsp/Timeseries/Data/dmi.had.\linebreak long.data, and information on Indian Ocean tropical cyclones from: https://www.metoc.navy.mil/jtwc/jtwc.html?best-tracks.

\bibliographystyle{ametsocV6}
\bibliography{Article_refs}

\begin{thebibliography}{33}
\providecommand{\natexlab}[1]{#1}
\providecommand{\url}[1]{\texttt{#1}}
\renewcommand{\UrlFont}{\rmfamily}
\providecommand{\urlprefix}{URL }
\expandafter\ifx\csname urlstyle\endcsname\relax
  \providecommand{\doi}[1]{https://doi.org/\discretionary{}{}{}#1}\else
  \providecommand{\doi}{https://doi.org/\discretionary{}{}{}\begingroup
  \urlstyle{rm}\Url}\fi
\providecommand{\eprint}[2][]{\url{#2}}

\bibitem[{Behera et~al.(2006)Behera, Luo, Masson, Rao, Sakuma,, and
  Yamagata}]{Behera2006cgcm}
Behera, S.~K., J.~J. Luo, S.~Masson, S.~A. Rao, H.~Sakuma, and T.~Yamagata,
  2006: A {CGCM} study on the interaction between {IOD} and {ENSO}. \textit{J.\
  Climate}, \textbf{19~(9)}, 1688--1705.

\bibitem[{Bessafi and Wheeler(2006)Bessafi, and
  Wheeler}]{bessafi2006modulation}
Bessafi, M., and M.~C. Wheeler, 2006: Modulation of south {I}ndian {O}cean
  tropical cyclones by the {M}adden--{J}ulian oscillation and convectively
  coupled equatorial waves. \textit{Mon.\ Wea.\ Rev.}, \textbf{134~(2)},
  638--656.

\bibitem[{Cai et~al.(2011)Cai, Sullivan,, and Cowan}]{Cai2011interactions}
Cai, W., A.~Sullivan, and T.~Cowan, 2011: Interactions of {ENSO}, the {IOD},
  and the {SAM} in {CMIP3} models. \textit{J.\ Climate}, \textbf{24~(6)},
  1688--1704.

\bibitem[{Chang et~al.(2004)Chang, Wang, Ju,, and Li}]{chang2004relationship}
Chang, C., Z.~Wang, J.~Ju, and T.~Li, 2004: On the relationship between western
  maritime continent monsoon rainfall and {ENSO} during northern winter.
  \textit{J.\ Climate}, \textbf{17~(3)}, 665--672.

\bibitem[{Chu et~al.(2002)Chu, Sampson, Levine,, and Fukada}]{chuEA2002}
Chu, J.-H., C.~R. Sampson, A.~S. Levine, and E.~Fukada, 2002: The {J}oint
  {T}yphoon {W}arning {C}enter tropical cyclone best-tracks, 1945-2000.
  \textit{NRL Reference Number: NRL/MR/7540-02-16}.

\bibitem[{Ciesielski et~al.(2014)}]{CiesielskiEA2014}
Ciesielski, P.~E., and Coauthors, 2014: Quality controlled upper-air sounding
  dataset for {DYNAMO}/{CINDY}/{AMIE}: {D}evelopment and corrections.
  \textit{J.\ Atmos.\ Oceanic Technol.}, \textbf{31}, 741--764,
  \doi{10.1175/JTECH-D-13-00165.1}.

\bibitem[{Ferreira et~al.(1996)Ferreira, Schubert,, and
  Hack}]{ferreira1996dynamical}
Ferreira, R.~N., W.~H. Schubert, and J.~J. Hack, 1996: Dynamical aspects of
  twin tropical cyclones associated with the {M}adden-{J}ulian oscillation.
  \textit{J.\ Atmos.\ Sci.}, \textbf{53}, 1520--0469.

\bibitem[{Fine et~al.(2016)Fine, Johnson, Ciesielski,, and Taft}]{Fine2016role}
Fine, C.~M., R.~H. Johnson, P.~E. Ciesielski, and R.~K. Taft, 2016: The role of
  topographically induced vortices in tropical cyclone formation over the
  {I}ndian {O}cean. \textit{Mon.\ Wea.\ Rev.}, \textbf{144~(12)}, 4827--4847.

\bibitem[{Gill(1980)}]{Gill1980}
Gill, A.~E., 1980: Some simple solutions for heat-induced tropical circulation.
  \textit{Quart.\ J.\ Roy.\ Meteor.\ Soc.}, \textbf{106}, 447--462.

\bibitem[{Gottschalck et~al.(2013)Gottschalck, Roundy, III, Vintzileos,, and
  Zhang}]{GottschalckEA2013}
Gottschalck, J., P.~E. Roundy, C.~J.~S. III, A.~Vintzileos, and C.~Zhang, 2013:
  Large-scale atmospheric and oceanic conditions during the 2011-12 {DYNAMO}
  field campaign. \textit{Mon.\ Wea.\ Rev.}, \textbf{141}, 4173--4196,
  \doi{10.1175/MWR-D-13-00022.1}.

\bibitem[{Hall et~al.(2001)Hall, Matthews,, and Karoly}]{hall2001modulation}
Hall, J.~D., A.~J. Matthews, and D.~J. Karoly, 2001: The modulation of tropical
  cyclone activity in the {A}ustralian region by the {M}adden--{J}ulian
  oscillation. \textit{Mon.\ Wea.\ Rev.}, \textbf{129~(12)}, 2970--2982.

\bibitem[{Hendon and Salby(1994)Hendon, and Salby}]{hendon1994life}
Hendon, H.~H., and M.~L. Salby, 1994: The life cycle of the {M}adden--{J}ulian
  oscillation. \textit{J.\ Atmos.\ Sci.}, \textbf{51~(15)}, 2225--2237.

\bibitem[{Hersbach et~al.(2020)}]{hersbach2020}
Hersbach, H., and Coauthors, 2020: The {ERA}5 global reanalysis.
  \textit{Quart.\ J.\ Roy.\ Meteor.\ Soc.}, \textbf{146},
  \doi{10.1002/qj.3803}.

\bibitem[{Hodges(1995)}]{hodges1995feature}
Hodges, K., 1995: Feature tracking on the unit sphere. \textit{Mon.\ Wea.\
  Rev.}, \textbf{123~(12)}, 3458--3465.

\bibitem[{Hodges(1999)}]{hodges1999adaptive}
Hodges, K., 1999: Adaptive constraints for feature tracking. \textit{Mon.\
  Wea.\ Rev.}, \textbf{127~(6)}, 1362--1373.

\bibitem[{Johnson and Ciesielski(2013)Johnson, and
  Ciesielski}]{Johnson+Ciesielski2013}
Johnson, R.~H., and P.~E. Ciesielski, 2013: Structure and properties of
  {M}adden-{J}ulian {O}scillations deduced from {DYNAMO} sounding arrays.
  \textit{J.\ Atmos.\ Sci.}, \textbf{70}, 3157--3179,
  \doi{10.1175/JAS-D-13-065.1}.

\bibitem[{Kiladis et~al.(1994)Kiladis, Meehl,, and
  Weickmann}]{kiladis1994large}
Kiladis, G.~N., G.~A. Meehl, and K.~M. Weickmann, 1994: Large-scale circulation
  associated with westerly wind bursts and deep convection over the western
  equatorial pacific. \textit{J.\ Geophy.\ Res.: Atmos.}, \textbf{99~(D9)},
  18\,527--18\,544.

\bibitem[{Kiladis et~al.(2005)Kiladis, Straub,, and Haertel}]{KiladisEA2005}
Kiladis, G.~N., K.~H. Straub, and P.~T. Haertel, 2005: Zonal and vertical
  structure of the {M}adden-{J}ulian oscillation. \textit{J.\ Atmos.\ Sci.},
  \textbf{62}, 2790--2809.

\bibitem[{Kuettner(1967)}]{kuettner1967equatorial}
Kuettner, J., 1967: Equatorial double vortex-unique hydrodynamic role of
  {S}umatra in atmospheric developments over {I}ndian {O}cean. \textit{Bull.\
  Amer.\ Meteor.\ Soc.}, \textbf{48~(8)}, 637.

\bibitem[{Kuettner(1989)}]{kuettner1989easterly}
Kuettner, J., 1989: Easterly flow over the cross equatorial island of {S}umatra
  and its role in the formation of cyclone pairs over the {I}ndian {O}cean.
  \textit{Wetter Leben}, \textbf{41}, 47--55.

\bibitem[{Madden and Julian(1971)Madden, and Julian}]{Madden1971}
Madden, R.~A., and P.~R. Julian, 1971: Detection of a 40–50 day oscillation
  in the zonal wind in the tropical {P}acific. \textit{J.\ Atmos.\ Sci.},
  \textbf{28}, 702--708, \doi{10.1175/1520-0469(1971)028<0702:DOADOI>2.0.CO;2}.

\bibitem[{Maloney and Hartmann(2000)Maloney, and
  Hartmann}]{maloney2000modulation}
Maloney, E.~D., and D.~L. Hartmann, 2000: Modulation of eastern {N}orth
  {P}acific hurricanes by the {M}adden--{J}ulian oscillation. \textit{J.\
  Climate}, \textbf{13~(9)}, 1451--1460.

\bibitem[{Ramage(1968)}]{ramage1968role}
Ramage, C.~S., 1968: Role of a tropical “maritime continent” in the
  atmospheric circulation. \textit{Mon.\ Wea.\ Rev.}, \textbf{96~(6)},
  365--370.

\bibitem[{Rotunno and Smolarkiewicz(1991)Rotunno, and
  Smolarkiewicz}]{rotunno1991further}
Rotunno, R., and P.~K. Smolarkiewicz, 1991: Further results on lee vortices in
  low-{F}roude-number flow. \textit{J.\ Atmos.\ Sci.}, \textbf{48~(19)},
  2204--2211.

\bibitem[{Rui and Wang(1990)Rui, and Wang}]{rui1990development}
Rui, H., and B.~Wang, 1990: Development characteristics and dynamic structure
  of tropical intraseasonal convection anomalies. \textit{J.\ Atmos.\ Sci.},
  \textbf{47~(3)}, 357--379.

\bibitem[{Saji et~al.(1999)Saji, Goswami, Vinayachandran,, and
  Yamagata}]{saji1999dipole}
Saji, N., B.~N. Goswami, P.~Vinayachandran, and T.~Yamagata, 1999: A dipole
  mode in the tropical {I}ndian {O}cean. \textit{Nature}, \textbf{401~(6751)},
  360--363.

\bibitem[{Smith(1989)}]{smith1989mountain}
Smith, R.~B., 1989: Mountain-induced stagnation points in hydrostatic flow.
  \textit{Tellus A: Dyn.\ Meteor.\ and Ocean.}, \textbf{41~(3)}, 270--274.

\bibitem[{Smolarkiewicz and Rotunno(1989)Smolarkiewicz, and
  Rotunno}]{smolarkiewicz1989low}
Smolarkiewicz, P.~K., and R.~Rotunno, 1989: Low {F}roude number flow past
  three-dimensional obstacles. {P}art {I}: {B}aroclinically generated lee
  vortices. \textit{J.\ Atmos.\ Sci.}, \textbf{46~(8)}, 1154--1164.

\bibitem[{Wheeler and Hendon(2004)Wheeler, and Hendon}]{Wheeler+Hendon2004}
Wheeler, M.~C., and H.~H. Hendon, 2004: An all-season real-time multivariate
  {MJO} index: {D}evelopment of an index for monitoring and prediction.
  \textit{Mon.\ Wea.\ Rev.}, \textbf{132}, 1917--1932.

\bibitem[{Wolter and Timlin(2011)Wolter, and Timlin}]{Wolter2011nino}
Wolter, K., and M.~S. Timlin, 2011: El {N}i{\~n}o/{S}outhern {O}scillation
  behaviour since 1871 as diagnosed in an extended multivariate {ENSO} index
  ({MEI}.ext). \textit{Int.\ J.\ Climatol.}, \textbf{31~(7)}, 1074--1087.

\bibitem[{Yoneyama et~al.(2013)Yoneyama, Zhang,, and Long}]{YoneyamaEA2013}
Yoneyama, K., C.~Zhang, and C.~N. Long, 2013: Tracking pulses of the
  {M}adden-{J}ulian {O}scillation. \textit{Bull.\ Amer.\ Meteor.\ Soc.},
  \textbf{94}, 1871--1891, \doi{10.1175/BAMS-D-12-00157.1}.

\bibitem[{Zhang(2013)}]{zhang2013madden}
Zhang, C., 2013: Madden--{J}ulian oscillation: {B}ridging weather and climate.
  \textit{Bull.\ Amer.\ Meteor.\ Soc.}, \textbf{94~(12)}, 1849--1870.

\bibitem[{Zhang et~al.(2019)Zhang, Hoell, Perlwitz, Eischeid, Murray,
  Hoerling,, and Hamill}]{Zhang2019towards}
Zhang, T., A.~Hoell, J.~Perlwitz, J.~Eischeid, D.~Murray, M.~Hoerling, and
  T.~M. Hamill, 2019: Towards probabilistic multivariate {ENSO} monitoring.
  \textit{Geophys.\ Res.\ Lett.}, \textbf{46~(17-18)}, 10\,532--10\,540.

\end{thebibliography}

tar
\clearpage

%

%
%
%



\end{document}